\documentclass{iopart}

\pdfoutput=1

\usepackage{color}
\usepackage{graphicx,epsfig,wasysym}
\usepackage{bm}
\usepackage{amssymb}

\def\be{\begin{equation}}
\def\ee{\end{equation}}

\def\bea{\begin{eqnarray}}
\def\eea{\end{eqnarray}}
\def\nn{\nonumber\\}
\def\nm{\newmoon}
\def\fm{\fullmoon}

\def\a{\alpha}
\def\s{\sigma}

\begin{document}

\title[The entanglement entropy of  1D systems in continuous and homogenous space]
{The entanglement entropy of one-dimensional systems in continuous and homogenous space}

\author{
Pasquale Calabrese, Mihail Mintchev, and Ettore Vicari
}
 \address{Dipartimento di Fisica dell'Universit\`a di Pisa and INFN, Pisa, Italy}

\date{\today}

\begin{abstract}

We introduce a systematic framework to calculate the bipartite
entanglement entropy of a compact spatial subsystem in a
one-dimensional quantum gas which can be mapped into a noninteracting
fermion system.  We show that when working with a finite number of
particles $N$, the R\'enyi entanglement entropies grow as $\ln N$,
with a prefactor that is given by the central charge.  We apply this
novel technique to the ground state and to excited states of periodic
systems.  We also consider systems with boundaries.  We derive
universal formulas for the leading behavior and for subleading
corrections to the scaling.  The universality of the results allows us
to make predictions for the finite-size scaling forms of the
corrections to the scaling. 

\end{abstract}

\maketitle

\section{Introduction}

Entanglement is a fundamental phenomenon of quantum mechanics.  Much
theoretical work has focused on the entanglement properties of quantum
many-body systems, showing their importance to characterize the
many-body dynamics~\cite{rev}.  In particular, lots of studies have
been devoted to quantify the nontrivial connections between
different parts of an extended quantum system, by computing von
Neumann or R\'enyi entanglement entropies of the reduced density
matrix $\rho_A$ of a subsystem $A$.  
R\'enyi entanglement entropies are defined as 
\be
S_\a=\frac1{1-\a}\ln{\rm Tr}\,\rho_{A}^\a\, .
\label{Sndef}
\ee For $\a\to1$ this definition gives the most commonly used von
Neumann entropy $S_1=-\tr{}{\rho_A\ln\rho_A}$, while for $\a\to\infty$
is the logarithm of the largest eigenvalue of $\rho_A$ also known as
single copy entanglement \cite{sce}.

One of the most remarkable results is the universal behavior displayed
by the entanglement entropy at 1D conformal quantum critical points
(i.e. with dynamical critical exponent $z=1$), determined by the
central charge \cite{c-lec} of the underlying conformal field theory
(CFT) \cite{holzhey,vidalent,cc-04,cc-rev}.  For a partition of an
infinite 1D system into a finite piece $A$ of length $\ell$ and the
remainder, the R\'enyi entanglement entropies for $\ell$ much larger
than the short-distance cutoff $a$ are \be
S_{\a}=\frac{c}6\left(1+\frac1\a\right) \ln \frac{\ell}a +c_\a\,,
\label{criticalent}
\ee where $c$ is the central charge and $c_\a$ a non-universal
constant.  When $A$ is a finite interval of length $\ell$ in a
finite periodic system of length $L$, CFT predicts the universal
asymptotic scaling \cite{cc-04} \be
S_{\a}=\frac{c}6\left(1+\frac1\a\right) \ln \Big(\frac{L}{\pi
a}\sin\frac{\pi\ell}L \Big)+c_\a\,,
\label{Sfinite}
\ee where, remarkably $c_\a$ is the same non-universal constant in
Eq. (\ref{criticalent}).  For future reference, it is also important
to mention the result in a finite system of length $L$ with some
boundary conditions at its ends and for an interval of length $\ell$
starting from one of the two boundaries \cite{cc-04,zbfs-06,lsca-06}:
\be S_{\a}=\frac{c}{12}\left(1+\frac1\a\right) \ln \Big(\frac{2L}{\pi
a}\sin\frac{\pi\ell}L \Big)+\frac{c_\a}2+\ln g\,,
\label{SfiniteB}
\ee where again $c_\a$ is the same non-universal constant as above and
$\ln g$ is the universal boundary entropy of Affleck and Ludwig
\cite{al-91}.  All the R\'enyi entropies $S_\a$ are proper and
equivalent measure of entanglement in a pure state \cite{rev}, but the
knowledge of $S_\a$ for different $\a$ characterizes the full spectrum
of non-zero eigenvalues of $\rho_A$ (see e.g. \cite{cl-08}) providing
significantly more information on the entanglement than the solely
knowledge of the von Neumann entropy.

The CFT results reviewed above have been confirmed
in many spin chains and in 1D itinerant systems on the lattice 
(too many to be mentioned here, we remand the interested reader to the 
comprehensive reviews on the subject \cite{rev}).  
These studies have allowed a deeper understanding of the convergence and precision \cite{t-08} of 
1D simulation algorithms based on the so-called matrix product states~\cite{mps}.  
However, analogous results must also be valid for systems in
continuous space, and therefore directly derivable in continuous models.
Apart from the interest to describe trapped 1D gases experimentally realized with
cold atoms, the entanglement of continuous models is also instrumental
to develop 1D tensor network algorithms for gases, as the one proposed in~\cite{vc-10}.  
Despite of this fundamental interest, almost no
effort (with the exception of Refs.~\cite{Klich-06,bk11} and the orbital
partitioning in quantum Hall states~\cite{QH}) has been devoted 
to the spatial entanglement of gas models (that is distinguished
from the particle partitioning~\cite{mzs-09}).

In a previous short communication \cite{us} we introduced a systematic framework to tackle free fermion gases
in any external conditions for an arbitrarily large number of particles.  
The most general result of this investigation was that,
when dealing with a finite number of particles $N$, the 1D
entanglement entropy grows like $\ln N$, with a prefactor that again
is given by the central charge.  
In this formulation $N$ acts as an UV cutoff, representing a concrete alternative to the lattice.  
In this manuscript, we detail the calculations in Ref. \cite{us} for  homogenous 1D gases
and we report a series of new results about the leading and subleading corrections to their scaling 
behavior. The degree of universality of these results allows to us 
to make novel predictions for spin chains on some universal functions  describing the corrections to the scaling.
The determination of these functions  was left as open problem by previous lattice investigations \cite{ccen-10,ce-10,xa-11}.

\subsection{The model and its equivalence with others}
\label{intro2}

We consider a system of free spinless fermions in the continuum
interval $[0,L]$.  We work with a finite number of particles $N$.
Therefore all the quantities and in particular entanglement entropies
are finite since $N$ acts as a cutoff.  Appropriate boundary
conditions (BC) are imposed in order to have a discrete energy
spectrum.

Apart the per se interest, spinless free fermions are also equivalent
to other models of direct physical application.  The 1D Bose gas with
short-ranged repulsive interaction (i.e. the Lieb-Liniger model
\cite{LL-63}) with Hamiltonian
\begin{equation}
{\cal H}_N = -\sum_{j=1}^N \frac{\partial^2}{\partial x_j^2} + 
2 C \sum_{1 \leq j < l \leq N} \delta(x_j - x_l)\,,
\end{equation}
in the limit of strong interaction $C\to\infty$ (i.e. impenetrable
bosons, also known as Tonk-Girardeau gas) is exactly mapped to
spinless fermions \cite{TG} and the entanglement entropy of a single
interval in the two models do coincide, because the boson in an
interval are functions only of the fermions in the same interval (this
is not true anymore in the case of more disjoint intervals because of
the presence of a bosonization string, analogously to 
spin-chain models~\cite{atc-09,ip-09,fc-10}).  The properties of the Lieb-Liniger model
are described solely by the dimensionless parameter $\gamma=C L/N$
\cite{LL-63}, thus the Tonks-Girardeau limit describes the dilute
model (i.e. $N/L\ll 1$) for any value of $C$.

Another important model mappable to free fermions is the spin-$1/2$ XX
chain defined by the Hamiltonian \be H = - \sum_{l=0}^L {1\over 2} [
\sigma^x_l \sigma^x_{l+1} + \sigma^y_l \sigma^y_{l+1}] - h \sigma^z_l
,
\label{HXX}
\ee
where $\sigma_l^{x,y,z}$ are the Pauli matrices at site $l$. The
Jordan-Wigner transformation 
\be
c_l=\left(\prod_{m<l} \s^z_m\right)\frac{\s^x_l+i\s_l^y}2\,,
\ee
maps this model to the quadratic Hamiltonian of spinless fermions 
\be
H =
-\sum_{l=0}^L  c_l^\dagger c_{l+1}  + c_{l+1}^\dagger
c_{l} + 2h \left(c_l^\dagger c_{l}-\frac{1}{2}\right). 
\label{Hfermi}
\ee
Here $h$ represents the chemical potential for the spinless fermions
$c_l$, which satisfy canonical anti-commutation relations
$\{c_l,c^\dagger_m\}=\delta_{l,m}$. The Hamiltonian (\ref{Hfermi}) is
diagonal in momentum space and for $|h|<1$ the ground-state is a
 Fermi sea with filling 
\be
\nu=\frac{\arccos |h|}\pi.
\ee
Only for $|h|<1$ we are dealing with a gapless theory. 

The continuum limit of the Hamiltonian (\ref{Hfermi}) is then the
system of free-fermions we are considering in this paper and so all the
universal properties that do not depend on lattice regularization can
be obtained from the continuum model.  At this point, it is worth
discussing how to obtain the continuum limit in some details.  The
lattice model is formed by $L_{\rm lat}$ sites separated by the
lattice spacing $a$ (usually set to $1$ in all lattice studies).  $N$
particles populate the chain with filling $\nu=N/L_{\rm lat}$ and we
are interested in the entanglement entropy of $\ell_{\rm lat}$ sites.
The continuum limit is a system of $N$ free fermions in a box of
length $L$ and is obtained by sending $a\to0$, $\nu\to0$ keeping fixed
$\nu \ell_{\rm lat}$ equal to $\ell N/L$ in the continuum, where
$L=aL_{\rm lat}$ and $\ell=a \ell_{\rm lat}$.  This allows us to use the
CFT results to predict a priori some of the results we are going to
derive.  In Ref. \cite{jk-04}, Eq. (\ref{criticalent}) has been
derived for the XX chain, obtaining \be
S_\a^{XX}=\frac{c}6\left(1+\frac1\a\right) \ln \Big(\frac{\ell_{\rm
lat}}a 2\sin \pi \nu \Big)+E_\a\,,
\label{JK}
\ee
where also the non-universal constant is determined as
\be\fl
E_\a=
\left(1+\frac1\a\right)\int_0^\infty \frac{dt}t 
\left[\frac{1}{1-\a^{-2}}
\left(\frac1{\a\sinh t/\a}-\frac1{\sinh t}\right)
\frac1{\sinh t}-\frac{e^{-2t}}6\right]\, .
\label{cnp}
\ee Combining this exact result with the CFT prediction in
Eq. (\ref{Sfinite}), we get the asymptotic scaling behavior of the
entanglement entropies in finite XX chain \be S_{\a}^{XX}
=\frac{c}6\left(1+\frac1\a\right) \ln \Big[\frac{L_{\rm lat}}{\pi
a}\sin\Big(\frac{\pi\ell_{\rm lat}}{L_{\rm lat}}\Big) 2\sin \pi \nu
\Big]+E_\a\,.
\label{SfiniteJK}
\ee
Taking now the continuum limit $a\to0$, $\nu\to 0$ as explained above, we arrive at the prediction 
\be\fl
S_{\a}^{\rm cont} \equiv \lim_{a\to 0}\Big[
S_{\a}^{XX}-\frac16\Big(1+\frac1\a\Big)\ln (1/a)\Big]=\frac16\left(1+\frac1\a\right)\ln\left (2{N}\sin\pi \frac\ell{L}\right)+ E_\a \,,
\label{pred}
\ee 
 where the subtraction of the term proportional to $\ln a$ comes from the normalization of the reduced density matrix. 

Finally we quote the 1D Bose-Hubbard model described by the Hamiltonian
\be
H_{\rm BH} = {J\over 2}
\sum_{i} (b_{i+1}-b_i)^\dagger (b_{i+1}-b_i) + {U\over2} \sum_i n_i(n_i-1) ,
\label{bhmN}
\ee
where  $b_i$ are bosonic operators and $n_i\equiv b_i^\dagger b_i$ is the particle density operator.
The hard-core limit $U\to\infty$ of the Bose-Hubbard model implies that the
particle number $n_i$ per site is restricted to the values $n_i=0,1$, and so
in  this limit can be exactly mapped into a lattice model of spinless fermions. Clearly the continuum limit 
of the Bose-Hubbard model is nothing but the Lieb-Liniger gas introduced above.


\section{The method}  

We consider a system of $N$ non-interacting spinless fermions with discrete one-particle energy spectrum, 
such as a finite system or one confined by a proper external potential.  
The many body wave functions $\Psi(x_1,...,x_N)$ can be written in terms of the  one-particle eigenstates as
a Slater determinant
\be 
\Psi(x_1,...,x_N)=\frac1{\sqrt{N!}}{\rm det} [\phi_k(x_n)],
\ee 
where the normalized wave functions $\phi_k(x)$ represent the occupied single-particle energy levels. 
The ground state is obtained by filling the $N$ levels with lowest energies.
Thus, the ground-state two-point correlator is
\begin{equation}
C(x,y) \equiv \langle c^\dagger(x) c(y) \rangle = 
\sum_{k=1}^{N} \phi^*_k(x) \phi_k(y)\,, 
\label{cxy}
\end{equation} 
where $c(x)$ is the fermionic annihilation operator and the one-particle
eigenfunctions $\phi_k(x)$ are intended to be ordered according to their energies.
The reduced density matrix of a subsystem $A$ extending from $x_1$ to $x_2$ can be written as
\be
\rho_A \propto \exp \Big(- \int_{x_1}^{x_2} d y_1 dy_2 c^\dagger(y_1) {\cal H}(y_1,y_2) c(y_2)\Big)\,,
\label{rhoa}
\ee
where ${\cal H}=\ln [(1-C)/C]$ and the normalization constant is fixed requiring ${\rm Tr}\rho_A=1$.
This equation can be straightforwardly seen as the continuum limit of the formula for lattice free fermions \cite{p-lat,ep-rev}, but 
can also be  obtained following the standard derivation in Ref. \cite{p-lat} in path integral formalism. 
In the passage from lattice to continuum, the normalization factor in (\ref{rhoa}) depends explicitly on the lattice spacing $a$
and it is responsible for the subtraction of the term proportional to $\ln a$ in Eq. (\ref{pred}).

We want to compute the bipartite Renyi entanglement entropies defined as in Eq. (\ref{Sndef}) of the space interval $A$ in this fermion gas. 
For this purpose, we introduce the {\it Fredholm determinant}\footnote{A Fredholm determinant is the extension of the standard 
determinant to {\it continuous matrices}. 
Its simplest operative definition is through the generalization to continuous kernels
$K(x,y)$ of the standard identity for determinants of a finite matrix ${\mathbb M}$
\be
\fl \ln  {\rm det}\left[{\lambda {\mathbb I}-{{\mathbb M}}}\right] =-\sum_{k=1}^\infty { {\rm Tr} {\mathbb M}^k \over k \lambda^k }
\Longrightarrow
\ln  {\rm det}\left[\lambda \delta(x-y)- K(x,y)\right] =-\sum_{k=1}^\infty { {\rm Tr} K^k \over k \lambda^k }\,,
\label{FDdef}
\ee
where the traces are simply
\be
 {\rm Tr} K^n=\int dx_1 dx_2\dots dx_n K(x_1,x_2)K(x_2,x_3)\dots K(x_{n-1},x_n) K(x_n,x_1)\,.
\ee
}
\begin{equation}
D_A(\lambda) =
{\rm det}\left[\lambda \delta_A(x,y)- C_A(x,y)\right] \,,
\label{dl}
\end{equation}
where $C_A(x,y)$ is the restriction of $C(x,y)$ to the part at hand
from $x_1$ to $x_2$ that can be written as
$C_A(x,y)=\theta(x-x_1)\theta(x_2-x) C(x,y)
\theta(y-x_1)\theta(x_2-y)$ (or in matrix form $C_A=P_A C P_A$, where
$P_A$ is the projector on the interval $A$).  The same definition
holds for $\delta_A(x,y)=P_A\delta(x-y)P_A$.  Following the ideas for
the lattice model \cite{jk-04}, $D_A(\lambda)$ can be introduced in
such a way that it is a polynomial in $\lambda$ having as zeros the
$N$ eigenvalues of $C_A$.  The Gaussian form of $\rho_A$ in
Eq. (\ref{rhoa}) allows us to exploit the relation between the
eigenvalues of $\rho_A$ and $C_A$ to write
\begin{equation}
S_\a \equiv \frac{\ln {\rm Tr}\rho_A^\a}{1-\a} = 
 \oint \frac{d \lambda}{2\pi i}\, e_\a(\lambda) 
\frac{d \ln D_A(\lambda)}{d\lambda},
\label{snx}
\end{equation}
where the integration contour encircles the segment $[0,1]$, and
\begin{equation}
e_\a(\lambda) = {1\over 1-\a} 
\ln \left[{\lambda}^\a
+\left({1-\lambda}\right)^\a\right]\,.
\label{enx}
\end{equation}
For $\a\to 1$,  $e_1(\lambda)=-x\ln x-(1-x)\ln(1-x)$ and Eq. (\ref{snx}) reproduces the von Neumann definition.  
The integral representation (\ref{snx}) has been
already derived and used in the context of discrete chain models~\cite{jk-04}, 
thus involving the determinant of a standard matrix with
the lattice sites as indices.

The Fredholm determinant is turned into a standard one by introducing
the $N\times N$ {\em reduced overlap} matrix ${\mathbb A}$ (also considered in
Ref.~\cite{Klich-06}) with elements
\begin{equation}
{\mathbb A}_{nm} =  \int_{x_1}^{x_2} dz\, \phi_n^*(z) \phi_m(z),
\qquad n,m=1,...,N,
\label{aiodef}
\end{equation}
such that 
\bea 
\fl{\rm Tr}\, C_A^k &=& \int_{x_1}^{x_2} \Big(\prod_{j=1}^k dy_j \Big) C_A(y_1,y_2) \dots C_A(y_{k-1},y_k)C_A(y_k,y_1)=\nonumber \\\fl&=&
 \int_{x_1}^{x_2} \Big(\prod_{j=1}^k dy_j \Big) 
 \sum_{n_1=1}^N \phi_{n_1}^*(y_1) \phi_{n_1}(y_2) \dots \sum_{n_k=1}^N \phi_{n_k}^*(y_k) \phi_{n_k}(y_1)=\\ \fl&=&
 \sum_{n_1=1}^N\dots  \sum_{n_k=1}^N  \int_{x_1}^{x_2} d y_1 \phi_{n_1}^*(y_1) \phi_{n_k}(y_1) \dots
 \int_{x_1}^{x_2} d y_k \phi_{n_k}^*(y_k) \phi_{n_{k-1}}(y_k)
 = {\rm Tr}\,{\mathbb A}^k, \nonumber 
\eea
where from the first to the second line we use Eq. (\ref{cxy}).
Thus
\be
\fl \ln {D}_A(\lambda) =
-\sum_{k=1}^\infty { {\rm Tr} C_A^k \over k \lambda^k }=
-\sum_{k=1}^\infty { {\rm Tr} {\mathbb A}^k \over k \lambda^k }=
\ln  {\rm det}\left[{\lambda {\mathbb I}-{{\mathbb A}}}\right] = 
\sum_{m=1}^N \ln(\lambda-a_m)\,, 
\label{logdetn}
\ee 
where we use twice Eq. (\ref{FDdef}) and we denote with 
$a_m$ the eigenvalues of ${\mathbb A}$.
Inserting it into the integral (\ref{snx}), we obtain
\begin{equation}
S_\a(x_1,x_2) =  \oint  \frac{d \lambda}{2\pi i}
\sum_{m=1}^N {e_\a(\lambda)\over  \lambda  - a_m}  = 
\sum_{m=1}^N e_\a(a_m),
\label{snx2n}
\end{equation}
as a consequence of the residue theorem.

The matrix ${\mathbb A}$ is easily obtained for any non-interacting
model from the one-particle wave functions, as the definition
(\ref{aiodef}) shows.  Calculating the entanglement entropies is then
reduced to an $N\times N$ eigenvalue problem that can be easily solved
numerically and in some instances even analytically, as we are going
to show. 

\section{The ground-state of systems with periodic boundary conditions}

In a system of length $L$ with periodic boundary conditions (BC), the normalized one-particle
wave-functions are plane waves with integer wave numbers
\be 
\phi_{k}(x)=\frac{e^{2\pi i k x/L}}{\sqrt{L}}\,, \qquad  k\in{\mathbb Z}, 
\ee
and energy $E_k=2\pi^2k^2/L^2$.
For some physical problems, one has to impose anti-periodic BC for the fermion degrees of freedom (the appropriate BC 
can also depends on the parity of $N$)
and so the momentum is quantized in terms of semi-integer wave numbers. 
However, as long as we are interested in the entanglement entropy of a single interval, this does not change 
the final results because, as we shall see soon, the elements of the matrix $ {\mathbb A}$ depends only on the 
difference between momenta that are always integer.
Different results would be instead obtained for the entanglement of two disjoint intervals, similarly to what happens in 
CFT \cite{fps-09,cct-09} and lattice models \cite{atc-09,ip-09,fc-10}. 

The element of the overlap matrix between two one-particle eigenstates with wave number $k_1$ and $k_2$ is 
\bea
 {\mathbb A}_{k_1k_2}&=&\frac1L \int_{x_1}^{x_2} dz\, e^{-2\pi i k_1 x/L}e^{2\pi i k_2 x/L}\nonumber\\&=&
e^{\pi i (k_2-k_1) (x_1+x_2)/L} \frac{\sin \pi(k_1-k_2)(x_2-x_1)/L}{\pi (k_1-k_2)}\,.
\label{Apbcgs}
\eea
The elements of the matrix $ {\mathbb A}$ are not invariant under translation because of the explicit 
dependence on $x_1+x_2$ of the phase factor. 
However, the eigenvalues of  ${\mathbb A}$ do not depend on this phase factor (but the eigenvectors do) 
and so also the entanglement entropies are translational invariant, as they must be. 
Indeed, in the determinant of $\lambda {\mathbb I-A}$, for each column we can bring out of the determinant the factor 
$e^{-\pi i k_1 (x_1+x_2)/L}$ and for each row $e^{\pi i k_2 (x_1+x_2)/L}$. 
Since $k_1$ and $k_2$ run on the same set of  integers, the product of all these phases is 1,
regardless of the values of the $k$s. 
In the following we use this freedom to fix the phase factor to $1$ and denote $\ell=x_2-x_1$.

\begin{figure}[tbp]
\includegraphics[width=.8\textwidth]{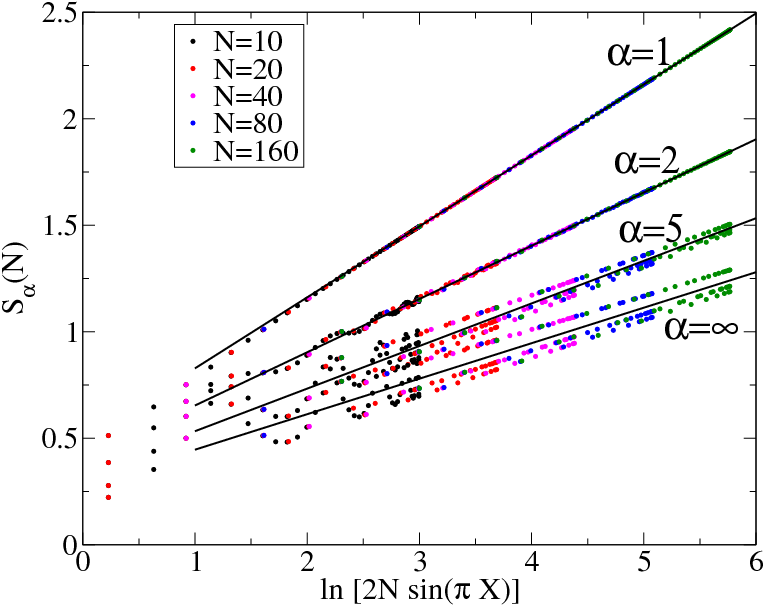}
\caption{ $S_\a(N)$ for $\a=1,2,5,\infty$ (from top to bottom) as function of $\ln[2N \sin(\pi X)]$ with $X=\ell/L$ calculated 
numerically for several values of $N=10,20,40,80,160$. The straight lines are the asymptotic predictions in Eq. (\ref{FHres}). 
The convergence is clear, although for $\a>1$ non-monotonic corrections to the asymptotic behavior are present.}
\label{fig:leading}
\end{figure}

The ground state of a fermion gas with $N$ particles is obtained by filling the $N$ $k$-modes with lowest energies.
In the case of odd $N$, this amounts to fill symmetrically the $N$ states with $|k|\leq (N-1)/2$ (the zero mode is clearly included).
For even $N$, there are two degenerate states obtained  from the $N-1$ ground state by filling the first available state
either on the right or on the left. This small difference between odd and even terms does not play any role 
because the elements of the matrix  ${\mathbb A}$ in Eq. (\ref{Apbcgs}) depend only on the differences between 
$k_1$ and $k_2$. 
Thus we can just start counting modes from the lowest $k$ one-particle occupied state and  
the resulting matrix ${\mathbb A}$ for a segment of length $\ell=x_2-x_1$ is  from Eq. (\ref{Apbcgs})
\be {\mathbb A}_{nm}= \frac{\sin \pi(n-m) \ell/L}{\pi (n-m)},
\qquad n,m=1,...,N\,.
\label{anmper}
\ee 
By inserting the $N$ eigenvalues of ${\mathbb A}$ into
Eq.~(\ref{snx2n}), we obtain the entanglement entropy in a system of $N$ particles.  
This is very easily done numerically as in Fig. \ref{fig:leading}.

\subsection{The leading behavior of the entanglement entropies}

We can also compute analytically the large $N$ behavior of the entanglement entropies.
Indeed we can write $\ln D_A=\ln \det {\mathbb G}$, where ${\mathbb G}\equiv \lambda {\mathbb I}-{\mathbb A}$ is  an $N\times N$
Toeplitz matrix. 
We can then use the Fisher-Hartwig conjecture~\cite{fh} to rigorously\footnote{Despite its name, the 
Fisher-Hartwig conjecture has been rigorously proven for the case at hand (see e.g. Ref. \cite{fh2}).} 
calculate the large $N$ behavior of $S_\a$.
However, going through all the technical complications of the Fisher-Hartwig conjecture is not needed, 
because we can exploit the results obtained for the very similar matrices of lattice free fermions on the infinite line.
In fact, for lattice models in the thermodynamic limit at fixed filling $\nu =N/L_{\rm lat}$, it has been found 
that the entanglement entropies of $\ell_{\rm lat}$ 
consecutive sites is given by Eq. (\ref{snx}) where $D_A(\lambda)$ is a standard $\ell_{\rm lat}\times\ell_{\rm lat}$ determinant 
with the correlation  matrix ${\bf C}$  given by \cite{ep-rev,gl-rev}
\be 
{\mathbb C}_{nm}^{\rm lat}= \frac{\sin \pi\nu(n-m) }{\pi (n-m)},
\qquad n,m=1,...,\ell_{\rm lat}\,.
\label{Clat}
\ee
It is evident that this matrix ${\mathbb C}_{nm}^{\rm lat}$ is the same as ${\mathbb A}$ in Eq. (\ref{anmper}) identifying 
$\nu$ with $ \ell/L$.
However, this is only a mathematical coincidence 
and  it will most probably not be true for interacting systems. 
Indeed, in Eq. (\ref{anmper}) we have a finite system and the indices are related to occupation modes of the $N$ particles in 
the full system. On the other hand, in Eq. (\ref{Clat}) we have an infinite lattice with filling $\nu$
and the indices refer to the lattice sites of the subsystem. 
Having established this equivalence between the two matrices, we can use the exact calculations 
in Ref. \cite{jk-04} (see also \cite{ce-10}), replace $\nu$ with $\ell/L$,
and obtain the asymptotic behavior of the desired entanglement entropies as  
\be
S_{\a} =\frac16\left(1+\frac1\a\right)\ln\left (2N\sin\pi \frac\ell{L}\right)+ E_\a 
+o(N^0). 
\label{FHres}
\ee 
Eq.~(\ref{FHres}) agrees with the CFT prediction for finite systems with periodic BC in Eq. (\ref{Sfinite}) and
represents an explicit analytic confirmation of this CFT result. It coincides with the scaling prediction in Eq. (\ref{pred}) from the 
lattice model, but here it has been derived from first principles.

Notice that we cannot recover the infinite volume limit from
Eq. (\ref{anmper}) because this limit must be taken at fixed ratio
$N/L$. If we naively take $L\to\infty$ in Eq. (\ref{anmper}) we get a
meaningless result, reflecting the non-commutation of the limits.
Oppositely, after computing the determinant as in
Eq. (\ref{FHres}), the infinite volume limit exists at finite density
$N/L$.

Fig.~\ref{fig:leading} shows a comparison with exact finite-$N$
calculations, for $\a=1,2,5,\infty$.  It is evident that (especially
for large $\a$) the data are affected by finite $N$ corrections that
are exactly calculated in the next subsection.

\subsection{Corrections to the asymptotic behavior and universal FSS in finite chains}

The above correspondence, between the determinants giving the
entanglement entropies of the continuos system and the ones for spin
chains, permits a quantitative description of the scaling correction
to the leading behavior in Eq. (\ref{FHres}) by exploiting the results
based on the {\it generalized} Fisher-Hartwig conjecture \cite{bt-91}
in Refs. \cite{ccen-10,ce-10}.  We introduce the differences between
the entanglement at finite $N$ and the asymptotic values $S_\a^{\rm
asy}(N)$ in Eq.  (\ref{FHres}) as \be d_\a(N)\equiv S_\a(N)-S_\a^{\rm
asy}(N)\,.  \ee

\begin{figure}[tbp]
\includegraphics[width=.8\textwidth]{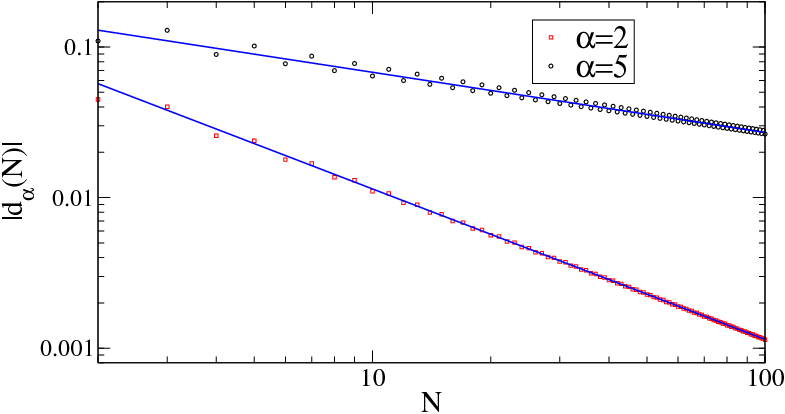}
\caption{Leading asymptotic correction to the R\'enyi entanglement entropies of half system $\ell=L/2$ for $N$ up to 100 and $\a=2,5$.
Straight lines correspond to the prediction (\ref{Sncorr}) that agrees with numerical data. 
Note that also subleading corrections oscillate and can be described by Eq. (\ref{fullresult}).  }
\label{fig:corral}
\end{figure}

We can again use the spin-chain results of Refs.~\cite{ccen-10,ce-10},
where a quantity analogous to $d_\a(N)$
was calculated at the leading order.  Using these
results and replacing $\nu$ with $\ell/L$ (in
\cite{jk-04,ccen-10,ce-10} $k_F=\pi\nu$ was used, but here we
prefer to use $\nu$ to avoid confusion with the Fermi momentum in the
continuous system), we obtain that the leading correction term is given by
\be
\fl d_\a(N)=
\frac{ 2 \cos(2 \pi \ell N/L) }{1-\a} (2N \sin \pi\ell/L)^{-2/\a}
\left[\frac{\Gamma(\frac{1}{2}+\frac{1}{2\a})}{\Gamma(\frac{1}{2}-\frac{1}{2\a})}\right]^2
+O\big(N^{-\min[4/\a,2]}\big).
\label{Sncorr}
\ee A check of the correctness of this expression is reported in
Fig. \ref{fig:corral}, where we report the absolute value of $d_\a(N)$
for the half system entanglement (i.e. $\ell/L=1/2$) for $N$ up to
100. The power law behavior is evident and the straight lines are
given by Eq. (\ref{Sncorr}) without any adjustable parameter.  These
corrections of the form $N^{-2/\a}$ correspond to the $\ell^{-2/\a}$
corrections found within conformal field theory \cite{cc-10}, that
have already been generalized to other situations, such as massive
field theories \cite{ccp-10}, confined systems \cite{cv-10},
disordered models \cite{fcm-11} and have been carefully checked
numerically in many different models \cite{ccen-10,x-10,xa-11}.

Subleading corrections to the scaling are visible in
Fig. \ref{fig:corral}, and for large values of $\a$ they have a
sizable effect.  These can be exactly calculated adapting the results
of Ref. \cite{ce-10} (based on generalized Fisher-Hartwig \cite{bt-91}
and random matrix theory \cite{fw-05}) and by replacing $\nu$ with
$\ell/L$.  The full result for $d_\a(\ell)$ up to order $N^{-3}$ can
be cast in the form \bea\fl
d_\a(N)&=&\frac{2}{\a-1}{\sum_{p,q=1}^\infty} (-1)^{p}
L_N^{-\frac{2p(2q-1)}{\a}}\big(Q_{q}\big)^p \left[ \frac{\cos(2 p\pi
N\ell/L )}{p} + \frac{A_q \sin(2p\pi N\ell/L)}{L_N} \right.  \nn
\fl&&\qquad\qquad\qquad\qquad\qquad\qquad \qquad\left.
+\frac{[B_{p,q}e^{2ip\pi N\ell/L}+{\rm h.c.}]}{L_N^2} \right] \nn\fl&&
+\frac{1}{L_N^2}\frac{\a+1}{285 \a^3}\left(15(3\a^2-7)+(49-\a^2)\sin^2
(\pi\ell/L)\right) + O\Big(L_N^{-3}\Big)\,,
\label{fullresult} 
\eea
where 
\bea
L_N&=&2 N \sin (\pi \ell/L)\,,
\label{Lk}\\
A_q&=&\left[1+3\left(\frac{2q-1}{\a}\right)^2\right]\cos (\pi\ell/L) \,,\\
Q_{q}&=&\left[\frac{\Gamma(\frac{1}{2}+\frac{2q-1}{2\a})}
{\Gamma(\frac12-\frac{2q-1}{2\a})}\right]^2\,,
\label{Qn}
\eea
\bea
B_{p,q}&=&\frac{2q-1}{6\a}\left[\big(5+7\frac{(2q-1)^2}{\a^2}\big)
\sin^2(\pi\ell/L)-15\big(\frac{(2q-1)^2}{\a^2}+1\big)\right]\nn
&&
-\frac{p}{4}\left[\big(1+3\frac{(2q-1)^2}{\a^2}\big)
\cos(\pi\ell/L)\right]^2.
\label{Bpq}
\eea
As for the spin chain, while Eq. (\ref{fullresult}) provides an infinite number of contributions, for a given fixed value of $\a$ 
only a finite number of them will be smaller than the leading neglected term, which is always of order $O(N^{-3})$. 
To be specific, in the cases $\a=2,3,10$ Eq.  (\ref{fullresult}) gives the leading $4,8,46$ terms in the asymptotic expansion of 
$d_\a(N)$ and hence the leading $6,10,48$ terms in the expansion of $S_\a(\ell)$. 
We do not report here all the terms which contribute for specific values of $\a$ that can be obtained from a simple adaptation of 
the results above or from Ref. \cite{ce-10}.

\begin{figure}[tbp]
\includegraphics[width=\textwidth]{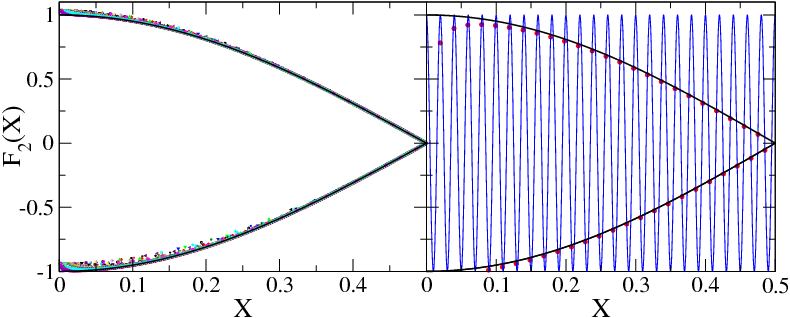}
\caption{Universal function for the correction to the scaling of the XX spin chain (lattice free fermion) $F_\a(X)$ in Eq. (\ref{Fal})
at zero magnetic field.
Left: Numerical results for spin chains with $\a=2$, an odd number of spins $17\leq L_{\rm }\leq 4623$, and all possible values of 
$\ell_{\rm lat}<L_{\rm lat}/2$. Data show perfect data collapse to the function $F_2(X)=\pm \cos\pi X$.
Right: numerical results for $L_{\rm lat}=101$ against our prediction $F_2(X)=\cos(2\pi \nu \ell_{\rm lat})$, 
showing graphically that the curve in the right panel is just the envelope obtained by sampling the oscillating function at 
integer values of $\ell_{\rm lat}$ (see the text for details). 
}
\label{fig:F2}
\end{figure}

A remarkable exact result that we obtain from the previous analysis is the universal finite size scaling (FSS)
form for finite XX spin chains with periodic BC. 
Indeed, as explained in section \ref{intro2}, the above result represents the continuum limit of 
the XX spin chain in a finite volume. 
For these spin chains, a universal FSS form has been observed in Ref. \cite{ccen-10}.
The quantity considered for the spin chain in Ref. \cite{ccen-10} is
\be\fl
F_\a(X)=\frac{d_\a(\ell_{\rm lat},L_{\rm lat})}
{\frac{2}{1-\a} \left(\frac{L_{\rm lat}}\pi \sin(\pi X) 2\sin \pi\nu\right)^{-2/\a}
\left[\frac{\Gamma(\frac{1}{2}+\frac{1}{2\a})}{\Gamma(\frac{1}{2}-\frac{1}{2\a})}\right]^2}
\,,\qquad {\rm with}\; X=\ell_{\rm lat}/L_{\rm lat}\,.
\label{Fal}
\ee
In Ref. \cite{ccen-10} chains with an {\it odd} number of spins $L_{\rm lat}$ have been considered. 
Fig. \ref{fig:F2} shows these results for $\a=2$ and magnetic field $h=0$.
The figure shows a perfect data collapse and somehow the correctness of the FSS ansatz. 
Already in Ref. \cite{ccen-10}, it was observed  that the scaling function is perfectly described 
by $F_2(X)=\pm \cos(\pi X)$,  as evident result from the figure (where it is impossible to distinguish the 
data from the conjecture). 
However, this is in apparent contradiction with our result, suggesting that a FSS form does not exist 
and the quantity $F_\a(X)$ in Eq. (\ref{Fal}) should instead be just a function of $\ell_{\rm lat}$
and in particular 
\be
F_\a(\ell_{\rm lat})=\cos(2\pi \nu \ell_{\rm lat})\,,
\label{Fanew}
\ee
where we changed the variable of $F_\a$ from $X$ to $\ell_{\rm lat}$.
To elucidate what is happening in the right panel of Fig. \ref{fig:F2}
we report the numerical data at $L_{\rm lat}=101$ in zero magnetic field against our new prediction Eq. (\ref{Fanew}).

Being $L_{\rm lat}$ odd, the ground state is not exactly at half-filling but at $\nu=(L_{\rm lat}-1)/2L_{\rm lat}$. 
In Fig. \ref{fig:F2} (right panel) we can observe that the prediction (\ref{Fanew}), i.e. the strongly oscillating curve,  
agrees with the numerical data,  apart some subleading corrections to the scaling.
The $\pm \cos(\pi X)$ form (also shown in the right panel) is nothing but the {\it sampling} of the curve at the integer 
values of $\ell_{\rm lat}$. Indeed from basic trigonometry we have
\bea\fl
\cos(2\pi \nu \ell_{\rm lat})&=& 
\cos\Big(\pi \frac{L_{\rm lat}-1}{L_{\rm lat}} \ell_{\rm lat}\Big)=
\cos\Big( \pi \ell_{\rm lat}- \pi \frac{\ell_{\rm lat}}{L_{\rm lat}}\Big)\\ \fl &=&
\cos\pi \ell_{\rm lat}\cos \pi \frac{\ell_{\rm lat}}{L_{\rm lat}}+\sin\pi \ell_{\rm lat}\sin \pi \frac{\ell_{\rm lat}}{L_{\rm lat}}
=(-1)^{\ell_{\rm lat}}\cos\pi X\,,\nonumber
\eea
where we used that $\ell_{\rm lat}$ is an integer. 
The final expression is exactly the phenomenological result conjectured in Ref. \cite{ccen-10} that we then prove and generalize here.
Indeed for even chains, always in zero field, Ref. \cite{ccen-10} proposed phenomenologically $F_\a(X)=\pm1$, which 
 corresponds to $\cos(2\pi\nu\ell_{\rm lat})$ i.e., for $\nu=1/2$, to $(-1)^{\ell_{\rm lat}}$. 
We checked Eq. (\ref{Fanew}) against other spin-chain results at different filling, always finding agreement.
Leading corrections to the scaling having a structure similar to Eq. (\ref{Fanew}) have been  conjectured in Ref. \cite{fc-11} 
for spin chains with open boundary conditions. 

\begin{figure}[tbp]
\includegraphics[width=\textwidth]{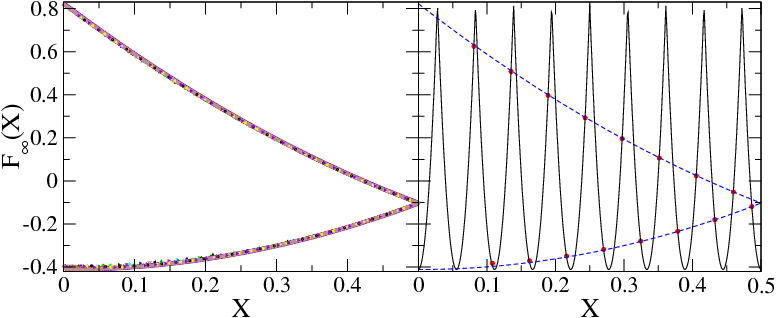}
\caption{Universal function $F_\infty(X)$ for the correction to the scaling of the single copy entanglement in the XX spin chain 
at zero-magnetic field.
Left: Numerical results for spin chains with an odd number of spins $17\leq L_{\rm }\leq 4623$ and all possible values of 
$\ell_{\rm lat}<L_{\rm lat}/2$. Data shows perfect data collapse to the function $F_\infty(X)$ in Eq. (\ref{infenv2}) shown as 
a continuous line, indistinguishable from data point.
Right: numerical results for $L_{\rm lat}=37$ against our prediction $F_\infty(X)$ in Eq. (\ref{infenv}). 
It is graphically evident that the curve on the left is the envelope obtained by sampling the oscillating functions at 
integer values of $\ell_{\rm lat}$. }
\label{fig:Finf}
\end{figure}

\subsubsection{The single copy entanglement: $\a\to\infty$.}
For $\a\to\infty$, the R\'enyi entanglement entropy gives the logarithm of the maximum eigenvalue of $\rho_A$
also known as single copy entanglement \cite{sce}.
It is not possible to obtain the result at $\a\to\infty$ from the general form in Eq. (\ref{fullresult}) because all the 
corrections of the form $N^{-2/\a}$ resum. 
Again instead of re-doing all the calculation to resum these terms, we exploit the correspondence with 
the infinite spin chain, and simply obtain the final result from Ref. \cite{ce-10} substituting $\nu$ with $\ell/L$.
After straightforward algebra we obtain
\be\fl
d_\infty(N)=\frac1{4\ln [b N\sin (\pi \ell/N)]}\left[{\rm Li}_2(-e^{i2 \pi\ell N/L })+{\rm Li}_2(-e^{-i2 \pi\ell N/L})\right],
\label{Finfi}
\ee
with $b=\exp(-\Psi(1/2))\approx 7.12429$.
We have checked these results against exact numerical computation that we do not report here.

It is interesting also in this case to explore the consequences of this result for finite spin chains, on the same lines as above
for the finite $\a$ results. 
In Ref. \cite{ccen-10}, for chains with an odd number of spins, it has been shown that the data for several choices of 
$L_{\rm lat}$ and $\ell_{\rm lat}$ collapse on a single master curve if plotted as
\be\fl
F_\infty(X)=\frac{d_\infty(\ell_{\rm lat},L_{\rm lat})}
{\Big[\ln \Big(2b \frac{L_{\rm lat}}{\pi} \sin(\pi X) \sin(\pi\nu) \Big)\Big]^{-1}}
\,,\qquad {\rm with}\; X=\ell_{\rm lat}/L_{\rm lat}\,.
\label{Finf}
\ee
Numerical data showing this collapse (analogous to those in Ref. \cite{ccen-10}) are reported in Fig. \ref{fig:Finf} (left) for zero 
magnetic field and odd $L_{\rm lat}$.
Oppositely to the case for finite $\a$, the shape of this curve was too complicated to be guessed in Ref. \cite{ccen-10}.
As before, assuming universality in the FSS towards the continuum limit, we predict
\be
F_\infty(\ell_{\rm lat})=\frac14 \left[{\rm Li}_2(-e^{i2 \pi\nu\ell_{\rm lat}  })+{\rm Li}_2(-e^{-i2 \pi\nu\ell_{\rm lat} })\right].
\label{infenv}
\ee
For a small chain with $L_{\rm lat}=37$, the above curve is reported in the right panel of Fig. \ref{fig:Finf} and it shows 
high frequency oscillations, but perfectly coincides with the exact lattice calculations (apart small subleading corrections).
As for finite $\a$, the smooth result obtained for chains of different length  (reported in the left panel)
is a consequences of the sampling at integer $\ell_{\rm lat}$. Using the property of the ${\rm Li}_2(y)$ function,
the envelope in the left panel of Fig. \ref{fig:Finf} is
\be
F_\infty(X)=\frac{(-1)^{\ell_{\rm lat}}}4 \left[{\rm Li}_2(-e^{i \pi X })+{\rm Li}_2(-e^{-i \pi X })\right],
\label{infenv2}
\ee
that is also shown in both panels, but in the left one is indistinguishable from data points. 

\subsubsection{The von Neumann entanglement entropy: $\a=1$.}
\begin{figure}[tbp]
\includegraphics[width=\textwidth]{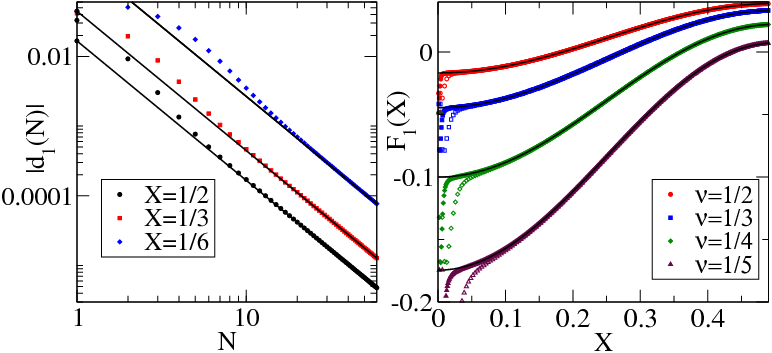}
\caption{Left: Leading correction to the scaling for the von Neumann entanglement entropy for the free-fermion gas 
for several length of the subinterval $\ell=X L$ as function of $N$ up to $60$.
The straight lines are the asymptotic prediction in Eq. (\ref{vN}).
Right: Universal function $F_1(X)$ in Eq. (\ref{vNF}) for the correction to the scaling of the entanglement entropy $S_1$ in the 
XX spin chain at zero-magnetic field as function of $X=\ell_{\rm lat}/L_{\rm lat}$.
Data are reported for two values of $L_{\rm lat}$ of order of 1000 and 300 which are compatible with the corresponding filling 
$\nu=1/2,1/3,1/4,1/5$. 
The continuous lines (indistinguishable from data points for not too small $X$) are the conjecture in Eq. (\ref{F1conj}).}
\label{fig:F1}
\end{figure}

For $\a=1$, the leading correction in Eq. (\ref{Sncorr}) is vanishing.
The actual calculation requires a complicated mapping with random matrix theory, but the final result, 
correct at $O(N^{-3})$, can be read from Eq. (\ref{fullresult})  
specialized to $\a=1$:
\be\fl
d_1(N)=-\frac1{12N^2}\left(\frac15+(\cot \pi\ell/L)^2\right)=
-\frac1{12N^2}\left(\frac1{(\sin \pi\ell/L)^2}-\frac45\right)\,.
\label{vN}
\ee
Fig. \ref{fig:F1} (left panel) reports the exact numerical computation for the entanglement
entropy for various values of $\ell/N$, showing perfect agreement with the asymptotic formula.

We consider again the XX spin chain for which the corrections to the scaling for the von Neumann entropy 
have not been considered quantitatively in finite systems. 
For infinite systems Ref. \cite{ce-10} reports the exact result
\be
d_1(\ell_{\rm lat})=-\frac1{12\ell_{\rm lat}^2}\left[\frac15+(\cot \pi \nu)^2\right]\,.
\label{vNl}
\ee
It should be pointed out that this term is an ``analytical correction'' to the scaling, i.e. it is not due to the insertion of 
an irrelevant operator. For this reason, its finite-size scaling cannot be obtained simply by replacing the distance $\ell_{\rm lat}$
with the chord length, as done for $d_\a$ at finite $N$.

For finite systems, we expect the FSS form
\be
d_1(\ell_{\rm lat},L_{\rm lat} )=\frac1{ 
\Big(\frac{L_{\rm lat}}\pi \sin(\pi\ell_{\rm lat}/L_{\rm lat}  )\Big)^2}F_1(\ell_{\rm lat}/L_{\rm lat} )\,,
\label{vNF}
\ee
where $F_1(X)$ is an unknown function with $F_1(0)$ fixed by Eq. (\ref{vNl}).
However,  by looking at its continuum limit, it is reasonable to propose the FSS ansatz
$F_1(X)= A +B \sin^2\pi X$. 
The constant $A$ can be fixed by requiring that $F_1(0)$ is given by Eq. (\ref{vNl}).
The constant $B$ can be fixed with the numerical data (e.g. by the scaling at $X=1/2$).
After a careful analysis, we conjecture the FSS scaling function
\be
F_1(X)= -\frac1{12}\left(\frac15+\cot^2 \pi \nu\right)+ \frac1{18}\left[1 + \frac65 \cot^2\pi\nu \right]\sin^2(\pi X) \,.
\label{F1conj}
\ee
By construction, this form reproduces Eq. (\ref{vNl}) for $X=0$. 
In the continuum limit, i.e. $\nu\to0$, it reproduces Eq. (\ref{vN}), as an highly  non-trivial check.
A stringent test of its correctness  is provided by numerical data. 
These are reported for few different values of $\nu$ and for  large values of $L_{\rm lat}$ in Fig. \ref{fig:F1} right panel, 
showing a perfect agreement.

\section{Excited states in periodic chains}

We now turn our attention to excited states that  can be easily treated in the formalism we introduced. 
Indeed, the only change compared to the ground state is in Eq.~(\ref{cxy}) where we have to  
sum over the occupied one-particle levels.
It is convenient to have a simple graphical representation of the many-body states.
This can be easily done by representing each single particle state with a circle and filling in black the occupied ones 
and leaving empty the others. For example, the ground state is
\be
\cdots \fm\fm \underbrace{\nm\cdots \nm\underline{\nm}\nm\cdots \nm}_{N}\fm\fm \cdots
\label{GSg}
\ee
where the underlined circle represents the zero momentum mode. 
When working at fixed number of particles $N$, excited states are obtained from the ground states just by 
moving black circles to empty white ones. 

The entanglement of excited states has been already considered few times in the literature, 
but only in the context of discrete lattice models. 
In \cite{as-08} it was shown that the negativity (which is a different measure of entanglement, related to some R\'enyi entropies \cite{rev})
 shows a universal scaling in critical spin systems. 
In Ref. \cite{afc-09}, on the bases of Toeplitz matrix arguments for the XX spin chain and by exact calculations for the anisotropic
Heisenberg one, it has been shown that only a small subclass of excited states 
can exhibit a {\it universal} logarithmic divergence with the
 subsystem size $\ell$, while most of the states strongly violates the area law and 
their entanglement entropies increase linearly with $\ell$, with non-universal prefactor.
The states providing universal scaling are those where there is a finite (and possibly small) number 
of sets of one-particle states occupied sequentially, as e.g.  
\be
\cdots \fm\fm\nm \nm\nm \fm\fm\fm \nm\nm\nm\nm \nm\fm\fm \cdots
\ee
and the locations of these blocks of states is not essential. 
This set of  states includes all low-lying excited states. 
In Ref. \cite{abs-11} it has been shown that the entanglement entropies of low-lying excited states
display a universal finite size scaling that is different from the one in the ground state of Eq. (\ref{Sfinite}).
These can be calculated by means of CFT, because low-lying states in CFT language are 
described by the action of a scaling  operator on the ground state. 
The states that are obtained by applying a primary operator to the ground states are of particular importance. 
In this case, the R\'enyi entropies for integer $\a$ have been related to the correlation functions of these operators
in a $\a$-sheeted Riemann surface \cite{abs-11}.
We remand the interested reader to the original reference \cite{abs-11} and we limit to quote the main result 
\be\fl
{\rm Tr}\rho_A^\a= [{\rm Tr}\rho_A^\a]_{(GS)} F^{(\a)}_\Upsilon(\ell/L)=c_\a \left(\frac{L}\pi \sin(\pi\ell/L) \right)^{c/6 (\a-1/\a)}F^{(\a)}_\Upsilon(\ell/L) \,.
\ee
where $ [{\rm Tr}\rho_A^\a]_{(GS)}$ is the ground-state value. $F^{(\a)}_\Upsilon(X)$ is the 
universal scaling function depending on the operator $\Upsilon$ whose action on the ground state gives the desired 
excited state. In particular $F^{(\a)}_\Upsilon(0)=1$, i.e., in the thermodynamic limit, all these low-lying states have entropies degenerate
with the ground-state, in agreement with Ref. \cite{afc-09}.

Two sets of primary operators can be easily treated for a free boson compactified on a circle, which describes the 
thermodynamic limit of the free-fermion gas we are considering. 
First, the vertex operators $V(x)$ for which Ref. \cite{abs-11} reports $F^{(\a)}_V(X)=1$ (i.e., the entanglement entropies 
are the same as in the ground-state).
In the free-fermion gas, this corresponds to the excited states obtained by shifting the ground-state (\ref{GSg})
in momentum space, i.e. replacing all $k_i$ with $k_i+M$ with $M$ arbitrary integer number.
The matrix ${\mathbb A}$ is always given by Eq. (\ref{Apbcgs}), that depends only of the differences between the 
various momenta, and so it is exactly equal to the ground-state one, confirming the prediction $F^{(\a)}_V(X)=1$.

The other operator considered in Ref. \cite{abs-11} is $\Upsilon=i\partial \phi$, which has been found to have a non-trivial 
scaling function given by
\be
F^{(\a)}_\Upsilon(X)= (-1)^\a \left(\frac2\a\sin{(\pi X)}\right)^{2\a}\det {\mathbb H}\,,
\label{predabs}
\ee
 for integer $\a$. Here $ {\mathbb H}$ is a $2\a\times 2\a$ matrix with elements
\be\fl 
{\mathbb H}_{jk}=\cases{ \frac1{e^{iz_j}-e^{iz_k}}& if $j\neq k$\\ 0&if $j= k$}
\quad {\rm and} \quad 
z_j=\cases{\pi(2j-2+x)/\a& if $j\leq \a$\\ \pi(2j-2-x)/\a& if $j> \a$}\,.
 \ee
For $\a=2$, this reduces to  the simple expression 
\be F^{(2)}(X)= \frac{(7+\cos(2\pi X))^2}{64},\ee 
but for any other $\a>2$ the explicit 
formulas are too cumbersome to be reported in their full glory. 
It must be mentioned that the analytic continuation of $F^{(\a)}_\Upsilon(X)$ is not yet known, and so also the von Neumann 
entanglement entropy of this excited state is still unknown. 
However for small $X$ it has been found
\be
F^{(\a)}_\Upsilon(X)=1-\frac{(\pi X)^2}3 \left( \a-\a^{-1}\right)+O(X^3)\,,
\label{Fsmallx}
\ee
whose analytic continuation is obvious. 

\begin{figure}[t]
\includegraphics[width=\textwidth]{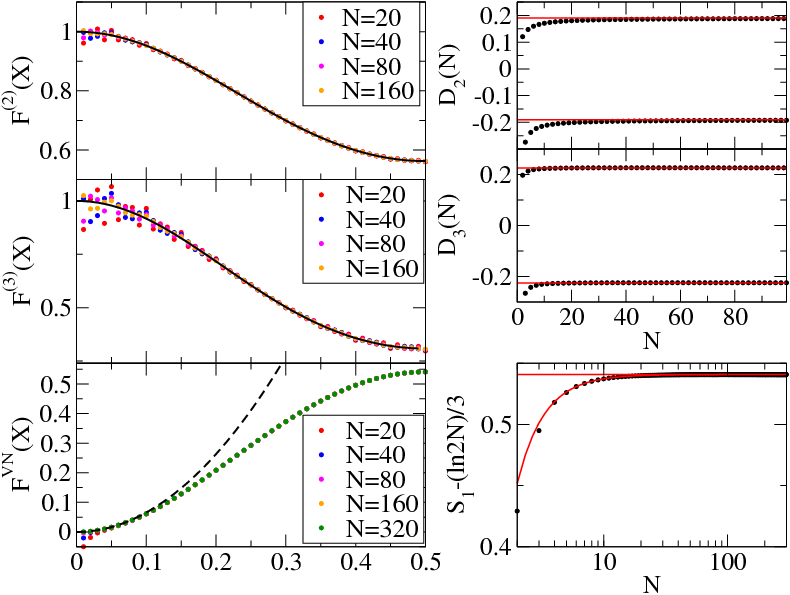}
\caption{Scaling functions for the entanglement R\'enyi entropies in the lowest energy particle-hole excited state of a periodic system 
in Eq. (\ref{ph-g}).
Left top panels report the scaling function $F^{(\a)}(X)$ for $\a=2,3$ as function of $X$, showing good agreement with the 
CFT prediction (\ref{predabs}) shown as continuous lines. 
The lower left panel reports the scaling function for the von Neumann entropy that for small $X$ agrees with the general expansion
(\ref{Fsmallx}), reported as a dashed line. 
On the right the universal corrections to the scaling are reported for the half-system entanglement 
showing the behavior $N^{-2/\a}$ for the leading corrections.
For $\a=1$, the corrections are monotonic (last panel) and effectively described by Eq. (\ref{s1excc}).}
\label{fig:exs}
\end{figure}

The excited state generated by the action of $i\partial\phi$ on the ground state is the 
particle-hole excitation obtained by moving one particle from the highest occupied level to the first available one,
i.e., graphically 
\be
\cdots \fm\fm \underbrace{\nm\nm\cdots \nm\underline{\nm}\nm\cdots \nm}_{N-1}\fm\nm\fm\fm \cdots
\label{ph-g}
\ee
The corresponding $N\times N$ matrix ${\mathbb A}$ has then the first $n-1$
rows and columns identical to the ground state Eq. (\ref{anmper}), but the last different, given by
\be {\mathbb A}_{Nm}= \cases{
\frac{\sin \pi(N+1-m) \ell/L}{\pi (N+1-m)}, &  $m=1,...,N-1$\\ \ell/L & $m=N$}\,,
\ee
and  $ {\mathbb A}_{mN}= {\mathbb A}_{Nm}$.
Despite only one row and one column differ from the ground-state, the matrix $ {\mathbb A}$ ceases to be a Toeplitz matrix
and (to the best of our knowledge) no analytic treatment is possible anymore.

We check the prediction in Eq. (\ref{predabs}) numerically.
In Fig.~\ref{fig:exs}, we report the numerical calculated scaling function
\be\fl
F^{(\a)}(X)\equiv\exp[(1-\a)({S_\a(N)- (1+1/\a)/6 \ln (2N\sin\pi X)-E_\a})]\,,
\ee
for several values of $N$  as a function of $X$ for $\a=2,3$.
It is evident that in the large $N$ limit  the CFT prediction (\ref{predabs}) is approached 
with small oscillating corrections to the scaling which are more pronounced for small $X$.
In order to shed some light on the analytic continuation at $\a\to1$, we also report (always in Fig. \ref{fig:exs})
the scaling function for the von Neumann entropy
\be
F^{VN}(X)\equiv S_1(N)- 1/3 \ln (2N\sin\pi X)-E_1\,.
\ee
As a difference with $F^{(\a)}$ with $\a\geq2$, the corrections to the scaling are much smaller, as for the entanglement in the 
ground-state. Unfortunately, as already stated, the analytic continuation to $\a\to1$ of Eq. (\ref{predabs}) for arbitrary $X$ is not yet known
and so the data in the figure cannot be contrasted to an exact prediction. However, such an analytic continuation is known 
for small $X$: from Eq. (\ref{Fsmallx}) we have $F^{VN}(X)=2(\pi X)^2 /3+O(X^3)$. This prediction is reported on top of the
numerical data and they agree perfectly up to $X\sim0.1$.

Having established the leading asymptotic behavior, we move our interest to the leading corrections to the scaling. 
We find numerically that the corrections have the same exponents as in the ground-state, i.e. they decay with 
$\a$ dependent power-law $N^{-2/\a}$. In order to show this, we report in Fig. \ref{fig:exs} (left panels)
the quantity
\be
D_\a(N)=[S_\a(N)-S_\a^{\rm asy}(N)] N^{2/\a}\,,
\ee
for the half-system entanglement. The data clearly show the behavior for the corrections of the form $(-1)^N N^{-2/\a}$ for $\a>1$.
We found numerically that $D_2\sim (-1)^N 0.19039\dots$ and $D_3\sim (-1)^N0.225\dots$. 
These non-universal amplitude are  different 
from the ones found for the ground state and we have been not able  neither to calculate nor to guess  their $\a$ dependence.  
We checked that for general $\ell/L$, the corrections are of the form  
$\cos(2\pi N\ell/L) N^{-2/\a}$ as for the ground state. 
Furthermore subleading corrections seem to have the same power structure as in the ground-state (cf.  
Eq. (\ref{fullresult})). 

The von Neumann entropy at $\a=1$ requires a separate analysis. Indeed, as the last panel of Fig. \ref{fig:exs} shows, the
correction to the scaling  are monotonic. However in this case we do not know exactly the constant term in the leading behavior. 
An accurate numerical analysis for the half-system entanglement  is consistent with the behavior
\be
S_1(N)= \frac13\log 2N+ y_1+\frac{y_2}{N^2}\,,
\label{s1excc}
\ee 
with $y_1=0.540726\dots$ and $y_2=0.35\dots$. These numerical data have been obtained by fitting data for $N>100,150,200$ and 
keeping under control the stability of the fit. 
Although these fitting parameters have been extracted from asymptotic large $N$, Fig. \ref{fig:exs} shows that the fit describes 
very accurately the data down to $N\sim3$.

\section{Systems with boundaries}

\subsection{Hard-wall boundaries}

We now consider a gas of
spinless fermions confined in the interval $[0,L]$ by a hard-wall
potential, i.e. the gas density vanishes outside the interval $x\notin [0,L]$ and the 
boundary condition is that the wave-function vanishes at the boundaries (Dirichlet BC). 
The one-particle wave functions are
\be
 \phi_k(x) =  {\sqrt{\frac2L}}\sin\left[{\pi}k\frac{x}L\right],
\qquad k=1,2,...,
\label{eq:pinfeig}
\ee
with energies $E_k = {\pi^2} k^2/2L^2$.

\begin{figure}[t]
\includegraphics[width=\textwidth]{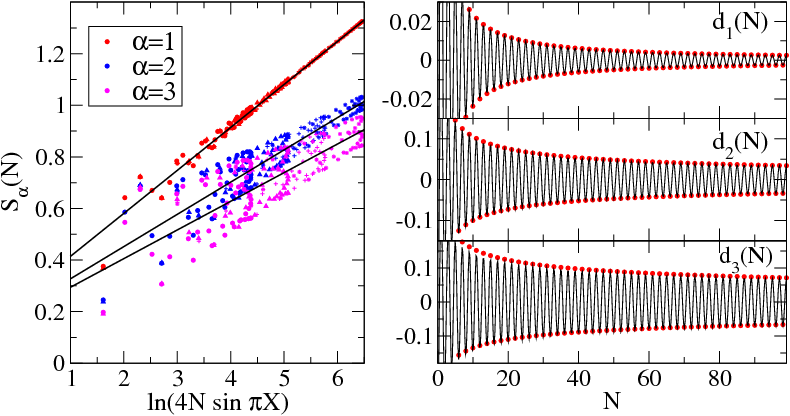}
\caption{On the left we report the R\'enyi entanglement entropies $S_\a$ for the free-fermion gas confined by hard wall potential
 for $\a=1,2,3$  (from top to bottom) as function of $\ln(4N\sin\pi X)$ with $X=\ell/L$. 
 For the larger values of $\a$,  oscillating corrections to the scaling obscure 
 the leading asymptotic behavior Eq. (\ref{FHres2}) represented as straight lines. 
 These oscillating corrections are quantitatively described by Eq. (\ref{daop}) that is checked in the right panel for the same 
 values of $\a$ for the half-system entanglement ($\ell=L/2$).}
\label{fig:op}
\end{figure}

\subsubsection{An interval starting from the boundary.}
The elements of the overlap matrix (cf. Eq. (\ref{aiodef})) between two one-particle eigenstates $n$ and $m$ have a particularly simple
form for an interval starting from the boundaries, i.e.   $A=[0,\ell]$. In fact we have 
\be\fl {\mathbb A_{nm}}=\int_{0}^{\ell} dz\, \phi_n^*(z) \phi_m(z)={\mathbb B}_{nm}(\ell)\equiv 
{\sin[\pi(n-m)\ell/L] \over \pi(n-m)} - {\sin[\pi(n+m)\ell/L] \over \pi(n+m)}.
\label{Snm}
\ee 
with $n,m=1,...,N$.

As for the periodic case, the matrix ${\mathbb A}$ above is exactly the same as the correlation matrix ${\mathbb C}^{\rm lat}$
of an infinite lattice with $\nu$ replaced by $\ell/N$.
This has been considered in Ref. \cite{fc-11} where, using a recent generalization of
the Fisher-Hartwig conjecture to Toeplitz+Henkel 
matrices~\cite{idk-09}, the asymptotic behavior of the entanglement entropies for the lattice model have been calculated exactly.  
Exploiting the equivalence between the two problems (i.e. replacing $\nu$ with $\ell/N$ in Ref. \cite{fc-11}) we
easily obtain for the asymptotic behavior of the entanglement entropies 
\be
S_{\a} =\frac1{12}\Big(1+\frac1\a\Big)\ln[2(2N+1)\sin\pi X]+ \frac{E_\a}2 + \cdots\,,
\label{FHres2}
\ee 
where $E_\a$ is defined in Eq. (\ref{cnp}).
Notice that this result agrees with the general CFT prediction in Eq. (\ref{SfiniteB}) with $\ln g=0$ 
that is a well-known result for open boundary conditions \cite{al-91}.

A comparison of the finite-$N$ results with Eq.~(\ref{FHres2}) is shown in Fig.~\ref{fig:op}.  
It is evident that for any $\a$ there are corrections to the scaling oscillating with $N$.
These are of the order $O(N^{-1/\a})$ and can be deduced exactly from the analogy 
with the lattice model solved in Ref. \cite{fc-11}.
Defining 
\be
d_\a(N)=S_\a(N)-S_\a^{\rm asy}(N)\,,
\ee
we have from Ref. \cite{fc-11} and replacing $\nu$ with $\ell/L$
\be\fl
d_\a(N)= \frac{2\sin[\pi (2N+1)\ell/L]}{1-\a}
[2(2N+1)\sin \pi\ell/L]^{-1/\a}\frac{\Gamma(\frac{1}{2}+\frac{1}{2\a})}{\Gamma(\frac{1}{2}-\frac{1}{2\a})}\,.
\label{daop}
\ee
Fig. \ref{fig:op} (right panel) show these corrections for half-system entanglement entropy for $\a=1,2,3$.
Further corrections of the form $N^{-p/\a}$ with $p$ integer can be straightforwardly deduced from the analysis in Ref. \cite{fc-11}.
We mention that, as for the periodic case, these leading $N^{-1/\a}$ corrections correspond to the ones of the form 
$O(\ell^{-{1/\a}})$ found within CFT~\cite{cc-10,fc-11}.


\subsubsection{Generic interval.}
In the case of a subsystem consisting of generic interval $A=[x_1,x_2]$, the entanglement entropies require a different analysis.
The general formula for the matrix ${\mathbb A}$ is slightly more complicated:
\be 
{\mathbb A_{nm}}=\int_{x_1}^{x_2} dz\, \phi_n^*(z) \phi_m(z)={\mathbb B}_{nm}(x_2)-{\mathbb B}_{nm}(x_1),
\label{Snmx1x2}
\ee 
where the matrix ${\mathbb B}$ is defined in Eq. (\ref{Snm}).
The entanglement entropies  of the interval
$[x_1,x_2]$ can be computed by inserting its eigenvalues in Eq.~(\ref{snx2n}).  
This allows us to easily compute
$S_\alpha(N)$ up to large values of $N$, and compare its
behavior with the asymptotic CFT prediction
\begin{eqnarray}
\fl S_\alpha(N) = \frac{1}6\Big(1+\frac1\a \Big) \left\{ \ln 4N + E_\a  
+ {1\over 2} \ln \Big[ 
{\sin^2[\pi(y_2-y_1)/2] \sin(\pi y_1) \sin(\pi y_2)
\over \sin^2[\pi(y_2+y_1)/2]}\Big]\right\},
\label{gens1}
\end{eqnarray}
where $y_i=x_i/L$.
The proof of this equation is a straightforward CFT exercise that we report in \ref{app}.

\begin{figure}[tbp]
\includegraphics[width=.5\textwidth]{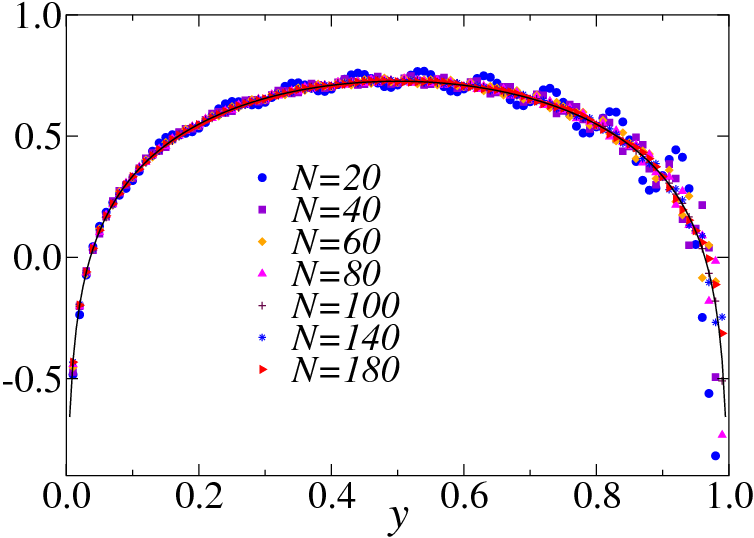}
\includegraphics[width=.5\textwidth]{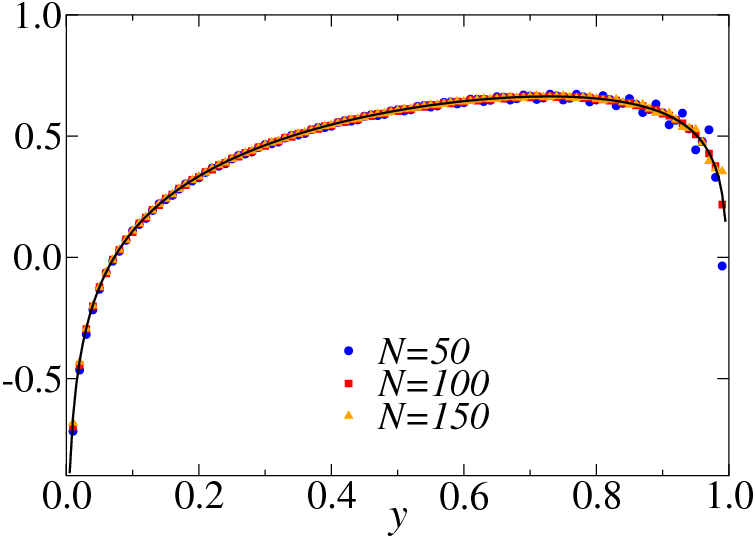}
\caption{(Color online) The von Neumann entanglement entropies $S_1$ 
of intervals $[x_1,x_2]$. 
In particular we consider the cases $x_1=L(1/2-y/2)$,
$x_2=L(1/2+y/2)$ (left) and $x_1=L/2$, $x_2=L(1/2+y/2)$ (right). 
We plot $S_1(N)-(\ln N)/6$ against $y$.  
In both figures the black full line shows the asymptotic behavior given by Eq.~(\ref{gens1}).  }
\label{dsbx}
\end{figure}

%

In particular, we considered a block of size $Ly$ centered at the
middle of the system, i.e., $y_2=1/2+y/2$ and $y_1=1/2-y/2$.  The data of the
entanglement entropy approach the asymptotic behavior predicted by
Eq.~(\ref{gens1}), i.e.,
\begin{equation}
S_\alpha(N) = \frac{1}6\Big(1+\frac1\a \Big) \left[ \ln 2N + E_\alpha +\ln\,\sin(\pi y) \right],
\label{appalbeh}
\end{equation}
as shown for $\a=1$ in Figs.~\ref{dsbx} (left panel).  
It can be seen numerically that the corrections to the scaling are of the order $O(N^{-1/\alpha})$.
It must be noted however that the $O(N^{-1/\alpha})$ convergence is nonuniform approaching the limits
$x\to 0$ and $x\to 1$, where the entanglement entropy trivially vanishes for any $N$.  
Simpler results are obtained also considering an interval starting from the center, i.e. taking
$x_2=1/2\pm y$ and $x_1=1/2$, for which
\begin{equation}
S_\alpha(N) =\frac{1}6\Big(1+\frac1\a \Big) \left\{ \ln 4N + E_\alpha
+ {1\over 2} \ln \left[ 
\cos(\pi y) \tan(\pi y/2)^2\right]
\right\},
\label{appalbeh2}
\end{equation}
see Fig.~\ref{dsbx} (right panel) for $\a=1$. We also checked the correctness of Eq. (\ref{gens1}) for other values of $\a$
and for different choices of $x_1$ and $x_2$, but the resulting figures are not very illuminating and  we do not report them.

\subsection{Neumann boundary conditions}

Another interesting situation arise when imposing Neumann boundary conditions on the fermionic wave-function, 
i.e. imposing that the derivative of the wave function vanishes at the two boundaries at $0$ and $L$. 
In this case, the normalized one-particle wave functions are 
\be
 \phi_k(x) = {\sqrt{\frac{2-\delta_{k,0}}L}}\cos\Big[\frac{\pi k x}L\Big],
\qquad k=0,1,...,
\label{eq:pinfeigneu}
\ee
with the same energy as for Dirichlet BC. As an important difference compared to Dirichlet BC, also the zero-mode 
with $k=0$ does contribute. 

The ${\mathbb A}$ matrix is readily calculated. It is an $N$-by-$N$ matrix with entries that are more easily written if we
count rows and columns with $n,m$ starting from $0$ and up to $N-1$ as for the modes above. 
For an interval of length $\ell$ starting from the boundary, straightforward calculations lead to 
\begin{eqnarray}
\fl&&{\mathbb A}_{nm} =  
\cases{\displaystyle {\sin[\pi(n-m)z] \over \pi(n-m)} +{\sin[\pi(n+m)z] \over \pi(n+m)}
 &  if  $n,m=1,...,N-1$, 
 \\ \displaystyle
\sqrt2  {\sin[\pi m z] \over \pi m} &if  $n=0$ and $m\neq 0$, \\ \displaystyle
 z&  if  $m,n=0$, 
}
\label{hlobneu}
\end{eqnarray}
and ${\mathbb A}_{0m} ={\mathbb A}_{m0} $. 
Note the plus sign between the two terms for $n,m\neq0$ and the zero-mode contribution, as a difference compared to Dirichlet BC.

\begin{figure}[t]
\includegraphics[width=\textwidth]{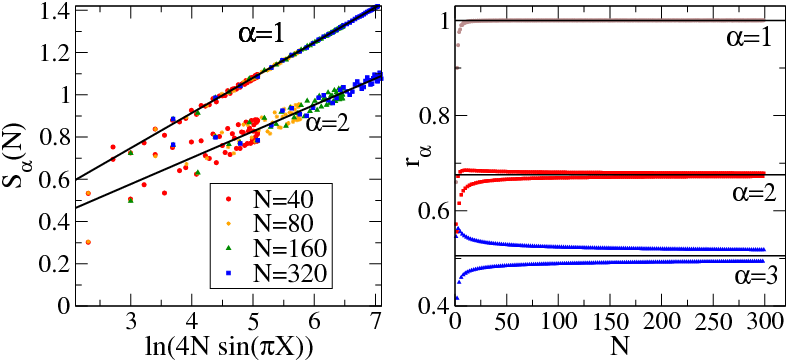}
\caption{On the left we report the R\'enyi entanglement entropies $S_\a$ for the free-fermion gas with Neumann BC 
 for $\a=1,2$  (from top to bottom) as function of $\ln(4N\sin\pi X)$ with $X=\ell/L$. 
 Oscillating corrections to the scaling are evident on top of 
 the leading asymptotic behavior Eq. (\ref{neumbc}) represented as straight lines. 
 These oscillating corrections are quantitatively described by Eq. (\ref{snalphanbcn}) that is checked in the right panel, 
  where we report the quantity $r_\a$ in Eq. (\ref{ral}) for $\a=1,2,3$ against the theoretical prediction from Eq. (\ref{snalphanbcn})
  for the half-system entanglement ($\ell=L/2$).}
\label{fig:neu}
\end{figure}

Because of the presence of the zero row and column, ${\mathbb A}$ is not of the form Toeplitz+Hankel as it is for Dirichlet BC.
Thus the recent generalizations of Fisher-Hartwig conjecture in Ref. \cite{idk-09} cannot be used. 
We then determine numerically  the matrix ${\mathbb A}$ for various $N$ and, through Eq. (\ref{snx2n}), 
we compute the R\'enyi entanglement entropy shown in Fig. \ref{fig:neu} (left panel).  
The analysis of their large-$N$ behavior gives
\begin{equation}
S_\a(N) = \frac1{12}\Big(1+\frac1\a\Big) \ln [2(2N-1) \sin(\pi \ell/L)]+ \frac{E_\a}2+\cdots\,,
\label{neumbc}
\end{equation}
shown as continuous lines in the figure. 
This form is consistent with the general CFT expectation in Eq. (\ref{SfiniteB}) with $g=1$. 
In order to avoid confusion with CFT literature, we  stress that, in this paper, we are considering Neumann and Dirichlet  BC
on the fermion degrees of freedom. These do {\it not} correspond to Dirichlet and Neumann BC on the bosonic field obtained from
the bosonization of the fermionic theory that instead are well known to have different $g$ function (see e.g. \cite{fsw-94}). 
It is known that they both correspond to Neumann conditions of the bosonized field and so it should be not a surprise
that the asymptotic behavior up to $O(N^0)$ is the same as Eq. (\ref{FHres2}).
Notice that we have included a $O(1/N)$ term in the leading behavior of the logarithm (i.e. the $-1$ in $(2N-1)$) that has the effect
to cancel the leading non-oscillating correction to the scaling. This was present also for Dirichlet BC, but it has {\it opposite} sign.
While before this was motivated by the mapping to the lattice model (cf. Ref. \cite{fc-11}), here we introduced it on a phenomenological
basis in order to describe the data (see below) and we do not have any mathematical explanation for it.

We now consider  the corrections in $N$ to the leading behavior, that are again consistent with the general scaling from 
CFT $O(N^{-1/\a})$. 
On the basis of the numerics, we guess exactly the first correction to the scaling, and we can write the entanglement entropies as 
\bea
\fl S_\a(N)&=&  \frac1{12}\Big(1+\frac1\a\Big) \ln [2(2N-1) \sin(\pi \ell/L)]+ \frac{E_\a}2+\nn\fl &&+ \frac{2\sin[\pi (2N-1)\ell/L]}{1-\a}
[2(2N-1)\sin \pi\ell/L]^{-1/\a}\frac{\Gamma(\frac{1}{2}+\frac{1}{2\a})}{\Gamma(\frac{1}{2}-\frac{1}{2\a})}\,.
\label{snalphanbcn}
\eea
In Fig. \ref{fig:neu} (right panel) we show the evidences for this scaling for the half-system entanglement ($\ell=L/2$). 
We report the quantity
\be
r_\a\equiv \frac{\left|S_\a(N) - \frac16 \ln (2(2 N-1)) - E_\a/2\right|}{[2(2N-1)]^{-1/\a}}\,,
\label{ral}
\ee
that for $\a=1,2,3$ approaches for large $N$ the value predicted from the ansatz (\ref{snalphanbcn})
$r_\a= \frac{2}{1-\a}\frac{\Gamma(\frac{1}{2}+\frac{1}{2\a})}{\Gamma(\frac{1}{2}-\frac{1}{2\a})}$.
For $\a=1$ the leading corrections are of order $1/N$.
The figure shows that the absolute value for even and odd $N$ coincide (but they have opposite signs).
This confirms that  choosing to parametrize the leading term with $(2N-1)$ in Eq. (\ref{neumbc}) 
cancels the $1/N$ non-oscillating corrections completely.
Some of the factors $(2N-1)$ in Eq. (\ref{snalphanbcn}) are subleading and 
are not explicitly tested by the numerical data presented.
They have been introduced from the analogy with Dirichlet BC results. 
Finally, we stress that we do not have any mathematical basis to justify Eq. (\ref{snalphanbcn}):
while its general structure can be inferred by CFT \cite{cc-10} because fermionic Neuman BC are in the same universality 
class as Dirichlet BC,  the {\it non-universal} amplitude of the correction has been guessed and conjectured by exploiting 
the analogy with Dirichlet BC and tested agains numerical data.

We also considered numerically other situations, such as 
other bipartitions of the systems. 
However, none of these results present particularly relevant or unexpected features  to be mentioned here.

\section{Conclusions}

In this manuscript we report the details about the computation of the entanglement entropies of continuous systems (gases)
which have been anticipated in the short communication \cite{us}.
The most important ingredient to write down  the entanglement entropies in terms of finite determinants is the use of the 
reduced overlap matrix in Eq. (\ref{aiodef}).
The calculation of the entanglement entropies is then mapped to the solution of an eigenvalue problem of an $N\times N$
matrix, with $N$ being the number of particles of the gas.

For the ground state of a periodic system we obtain the leading behavior in the form 
\be
S_{\a} =\frac16\left(1+\frac1\a\right)\ln\left (2N\sin\pi \frac\ell{L}\right)+ E_\a  +o(N^0), 
\ee 
while for a gas with Dirichlet or Neumann boundary conditions we find
\be
S_{\a} =\frac1{12}\left(1+\frac1\a\right)\ln\left (4N\sin\pi \frac\ell{L}\right)+ \frac{E_\a}2 +o(N^0), 
\ee 
both in agreement with CFT and scaling expectations, but they have been found here from first principles. 
We also derive the corresponding leading behavior for some classes of excited states. 

Furthermore, adapting to the problem at hand the results in Refs. \cite{ce-10,fc-10}, we calculate also subleading 
corrections. 
The universality of these formulas allowed us to infer the finite-size scaling forms for spin chains which are reported 
in Eq. (\ref{Fanew}) for $1<\a<\infty$, in Eq. (\ref{Finfi}) for $\a=\infty$,  and in Eq. (\ref{F1conj}) for $\a=1$.
The determination of these exact formulas were left as open problems from previous investigations.

Some other applications (such as to systems with defects, star graphs,  and to gases confined by an external potential both in and out 
of equilibrium) of this novel method 
have been already shortly presented in Ref. \cite{us}, but they will be detailed elsewhere. 
Other generalizations,  which we are currently investigating, concern the calculation of the entanglement 
for quadratic Hamiltonian which do not conserve the
fermion number (such as the continuum limit of the XY model), free gases in higher dimensions and different geometries. 
Finally, some non equilibrium situations such as local quantum quenches  (e.g., instantaneously turning 
on/off of a defect) can also be tackled within this framework. 
The asymptotic CFT results in several circumstances are known \cite{lq}, but analytic calculations for specific models are still missing.  
They may provide important insights in view of the recent proposals of using the full counting statistics after a quench as an experimental probe and a measure of entanglement \cite{kl-08}.

\section*{Acknowledgments}

We thank Maurizio Fagotti and Fabian Essler for helpful discussions. We  thank German Sierra and Miguel Ibanez Berganza 
for correspondence about Ref. \cite{abs-11}.

\appendix

\section{The CFT entanglement entropy of an arbitrary interval in a finite system with Dirichlet boundary conditions}
\label{app}

We provide here the CFT proof of Eq. (\ref{gens1}). 
In CFT the moments of the reduced density matrix can be written as correlation function of particular {\it twist} fields
that transform as primary operators under a conformal transformation \cite{cc-04,cc-rev,ccd-08}.
In particular, for the case of a finite interval between $x_1$ and $x_2$ in a boundary theory, we have (for integer $n$)
\be
{\rm Tr}\,\rho_A^n=\langle\Phi_n(x_1)\Phi_{-n}(x_2)\rangle\,,
\label{twist}
\ee
where $\Phi_n$ and $\Phi_{-n}$  transform as primary fields with dimensions $x_n=(c/24)(n-1/n)$.
 
This two-point function in the finite strip with ${\rm Re}(w)\in [0,L]$ can be obtained from its conformal mapping 
to the upper half plane (UHP) with ${\rm Im}(z)>0$.
The mapping and its inverse are
\be
iw=\frac{L}\pi\ln z\,,\qquad z=e^{i \pi w/L}\,.
\label{mapp}
\ee

Using the property that the twist fields for free fermions with open boundary conditions behaves like primary 
operators in a free bosonic theory \cite{fc-11}, the two-point function in the UHP can be read from Ref. \cite{cardy-84}
\be
\langle\Phi_n(z_1) \Phi_{-n}(z_2) \rangle_{\rm UHP}=c_n\left
(\frac{z_{1\bar2}z_{2\bar1}}{z_{12}z_{\bar1\bar2}z_{1\bar1}z_{2\bar2}}\right)^{2x_n}\,,
\label{bosUHP}
\ee
with $z_{ij}=|z_i-z_j|$ and $z_{\bar k}={\overline{z_k}}$ and $c_n$ an undetermined constant (we set the UV cut-off $a$ to $1$).
We stress that in general (i.e. for a theory that is not free bosonic) a universal function of the harmonic ratio 
 build with the four points $z_i$ and $\bar z_i$ multliplies the above formula \cite{fps-09,cct-09}.
Using then the conformal mapping (\ref{mapp}), we have
\bea\fl
z_{12} z_{\bar1\bar2}&=&(e^{i\pi w_1/L}-e^{i\pi w_2/L}) (e^{-i\pi w_1/L}-e^{-i\pi w_2/L})=4\sin^2\frac{\pi(w_1-w_2)}{2L}\,, \nn \fl 
|z_{1\bar1}|&=&2\sin\frac{\pi w_1}L\,, \qquad |z_{2\bar2}|=2\sin\frac{\pi w_2}L\,,\nn \fl
z_{1\bar2}z_{\bar1 2}&=&(e^{i\pi w_1/L}-e^{-i\pi w_2/L}) (e^{-i\pi w_1/L}-e^{i\pi w_2/L})=4\sin^2\frac{\pi(w_1+w_2)}{2L}\,. 
\eea
Thus
\bea\fl
\langle\Phi_n(r,\tau) \Phi_{-n}(0,\tau) \rangle_{\rm strip}&=&
|w'(z_1)|^{-x_n} |w'(z_2)|^{-x_n} \langle \Phi_n(z_1(w)) \Phi_{-n}(z_2(w))\rangle_{\rm UHP}=\nn \fl &=&
c_n\left(\frac{\pi^2}{4L^2} \frac{\sin^2\frac{\pi(w_1+w_2)}{2L}}{\sin^2\frac{\pi(w_1-w_2)}{2L} \sin\frac{\pi w_1}L\sin\frac{\pi w_2}L}\right)^{2x_n}\,.
\eea

Using Eqs. (\ref{twist}) and (\ref{Sndef}), performing the analytic continuation from $n$ to $\a$, we have
\be\fl
S_\a=\frac{c}{6}\Big(1+\frac1\a\Big)\Big[\ln \frac{2L}{\pi}+\frac12 \ln
\frac{\sin^2\frac{\pi(w_1-w_2)}{2L} \sin\frac{\pi w_1}L\sin\frac{\pi w_2}L}{\sin^2\frac{\pi(w_1+w_2)}{2L}}\Big]+E_\a\,,
\ee
where the constant $E_\a$ is given by Eq. (\ref{cnp}), according to the result for the single interval in a periodic systems. 
Assuming now the scaling hypothesis when working with finite number of particles, Eq. (\ref{gens1})
follows simply by replacing ${L}/\pi$ by $2N$ using the argument in Section \ref{intro2}.

\end{document}